\documentclass[aip,reprint]{revtex4-1}
\usepackage{amsmath,amssymb}
\usepackage{graphicx}
\usepackage{dcolumn}
\usepackage{bm}
\usepackage{subfigure}
\usepackage{pifont} 
\usepackage{siunitx}

\usepackage[utf8]{inputenc}
\usepackage[T1]{fontenc}
\usepackage{mathptmx}
\usepackage{etoolbox}
\usepackage{color}
\makeatletter
\def\@email#1#2{%
 \endgroup
 \patchcmd{\titleblock@produce}
  {\frontmatter@RRAPformat}
  {\frontmatter@RRAPformat{\produce@RRAP{*#1\href{mailto:#2}{#2}}}\frontmatter@RRAPformat}
  {}{}
}%
\makeatother
\begin{document}

\preprint{AIP/123-QED}

\title{Demonstration of High-Fidelity Gates in a Strongly Anharmonic with Long-Coherence C-Shunt Flux Qubit}
\author{Silu Zhao}
\thanks{These authors contributed equally to this work.}
\affiliation{
	Beijing National Laboratory for Condensed Matter Physics,\\
	Institute of Physics,Chinese Academy of Sciences,Beijing 100190,China
}
\affiliation{
School of Physical Sciences, University of Chinese Academy of Sciences, Beijing 100049, China
}
\author{Li Li}
\thanks{These authors contributed equally to this work.}
\affiliation{
	Beijing National Laboratory for Condensed Matter Physics,\\
	Institute of Physics,Chinese Academy of Sciences,Beijing 100190,China
}
\affiliation{
School of Physical Sciences, University of Chinese Academy of Sciences, Beijing 100049, China
}
\author{Weiping Yuan}
\thanks{These authors contributed equally to this work.}
\affiliation{
	Beijing National Laboratory for Condensed Matter Physics,\\
	Institute of Physics,Chinese Academy of Sciences,Beijing 100190,China
}
\affiliation{
School of Physical Sciences, University of Chinese Academy of Sciences, Beijing 100049, China
}
\author{Xinhui Ruan}
\affiliation{
Beijing National Laboratory for Condensed Matter Physics,\\
Institute of Physics,Chinese Academy of Sciences,Beijing 100190,China
}

\author{Jinzhe Wang}
\affiliation{
	Beijing National Laboratory for Condensed Matter Physics,\\
	Institute of Physics,Chinese Academy of Sciences,Beijing 100190,China
}
\affiliation{
School of Physical Sciences, University of Chinese Academy of Sciences, Beijing 100049, China
}
\author{Bingjie Chen}
\affiliation{
	Beijing National Laboratory for Condensed Matter Physics,\\
	Institute of Physics,Chinese Academy of Sciences,Beijing 100190,China
}
\affiliation{
School of Physical Sciences, University of Chinese Academy of Sciences, Beijing 100049, China
}

\author{Yunhao Shi}
\affiliation{
	Beijing National Laboratory for Condensed Matter Physics,\\
	Institute of Physics,Chinese Academy of Sciences,Beijing 100190,China
}

\author{Guihan Liang}
\affiliation{
	Beijing National Laboratory for Condensed Matter Physics,\\
	Institute of Physics,Chinese Academy of Sciences,Beijing 100190,China
}

\author{Shi Xiao}
\affiliation{
	Beijing National Laboratory for Condensed Matter Physics,\\
	Institute of Physics,Chinese Academy of Sciences,Beijing 100190,China
}

\author{Jiacheng Song}
\affiliation{
	Beijing National Laboratory for Condensed Matter Physics,\\
	Institute of Physics,Chinese Academy of Sciences,Beijing 100190,China
}
\affiliation{
School of Physical Sciences, University of Chinese Academy of Sciences, Beijing 100049, China
}
\author{Jinming Guo}
\affiliation{
	Beijing National Laboratory for Condensed Matter Physics,\\
	Institute of Physics,Chinese Academy of Sciences,Beijing 100190,China
}
\affiliation{
School of Physical Sciences, University of Chinese Academy of Sciences, Beijing 100049, China
}
\author{Xiaohui Song}
\affiliation{
	Beijing National Laboratory for Condensed Matter Physics,\\
	Institute of Physics,Chinese Academy of Sciences,Beijing 100190,China
}
\affiliation{
	Hefei National Laboratory, Hefei 230088, China
}

\author{Kai Xu}
\affiliation{
	Beijing National Laboratory for Condensed Matter Physics,\\
	Institute of Physics,Chinese Academy of Sciences,Beijing 100190,China
}
\affiliation{
	Hefei National Laboratory, Hefei 230088, China
}
\affiliation{
Beijing Academy of Quantum Information Sciences, Beijing 100193, China
}

\author{Heng Fan}
\email{hfan@iphy.ac.cn}
\affiliation{
	Beijing National Laboratory for Condensed Matter Physics,\\
	Institute of Physics,Chinese Academy of Sciences,Beijing 100190,China
}
\affiliation{
	School of Physical Sciences, University of Chinese Academy of Sciences, Beijing 100049, China
}
\affiliation{
	Hefei National Laboratory, Hefei 230088, China
}
\affiliation{
Beijing Academy of Quantum Information Sciences, Beijing 100193, China
}

\author{Zhongcheng Xiang}
\email{zcxiang@iphy.ac.cn}
\affiliation{
	Beijing National Laboratory for Condensed Matter Physics,\\
	Institute of Physics,Chinese Academy of Sciences,Beijing 100190,China
}
\affiliation{
	Hefei National Laboratory, Hefei 230088, China
}

\author{Dongning Zheng}
\email{dzheng@iphy.ac.cn}
\affiliation{
	Beijing National Laboratory for Condensed Matter Physics,\\
	Institute of Physics,Chinese Academy of Sciences,Beijing 100190,China
}
\affiliation{
	School of Physical Sciences, University of Chinese Academy of Sciences, Beijing 100049, China
}
\affiliation{
	Hefei National Laboratory, Hefei 230088, China
}
	
\date{\today}

\begin{abstract}
We demonstrate high-fidelity single-qubit gates on a C-shunt flux qubit that simultaneously combines a large anharmonicity ($\mathcal{A}/2\pi=848~\mathrm{MHz}$) with long relaxation time ($T_1 = 23~\mu\text{s}$). The large anharmonicity significantly suppresses leakage to higher energy levels, enabling fast and precise microwave control. Using DRAG pulses and randomized benchmarking, the qubit achieves gate fidelities exceeding 99.9\%, highlighting the capability of C-shunt flux qubits for robust and high-performance quantum operations. These results establish them as a promising platform for scalable quantum information processing.
\end{abstract}

\maketitle
Superconducting qubits are one of the leading platforms for realizing scalable quantum computation~\cite{clarke2008,dicarlo2009,Devoret2013,Barends2014,blais2021}. Among them, the transmon qubit has been widely adopted due to its long coherence times and relatively simple control~\cite{Koch2007,houck2008,barends2013b}, but its small anharmonicity limits the speed and fidelity of quantum gate operations, and also leads to frequency crowding in multi-qubit devices.~\cite{motzoi2009,schutjens2013,chen2016}. Flux qubits, on the other hand, naturally possess large anharmonicity and strong tunability~\cite{Mooij1999,Orlando1999,chiorescu2004,dmitriev2021}, but typically suffer from short coherence times and increased sensitivity to flux noise~\cite{Yoshihara2006}.

To overcome these challenges, hybrid designs that combine the advantages of both transmon and flux qubits have attracted increasing attention~\cite{manucharyan2009,yan2016,liu2023}. Prominent examples include the fluxonium qubit, which has recently demonstrated remarkable coherence properties and high-fidelity gate operations~\cite{nguyen2019,somoroff2023,ding2023a,ma2024,zhang2024,an2025}, although it often requires complex fabrication of Josephson junction arrays and operates at relatively low frequencies. In particular, the capacitively shunted (C-shunt) flux qubit provides a promising route to balance anharmonicity and coherence by introducing a large shunt capacitance~\cite{yan2016,abdurakhimov2019}. Such a design has the potential to support both long-lived quantum states and high-fidelity gate operations. Moreover, its relatively simple fabrication requirements make it an attractive candidate for scalable quantum processors. Previous studies have primarily focused on device design, spectroscopy, and coherence properties of this architecture. A detailed assessment of gate-level control performance is therefore important for establishing its viability for practical quantum information processing.

    \begin{figure}[t]
    \centering
    \includegraphics[width=\columnwidth]{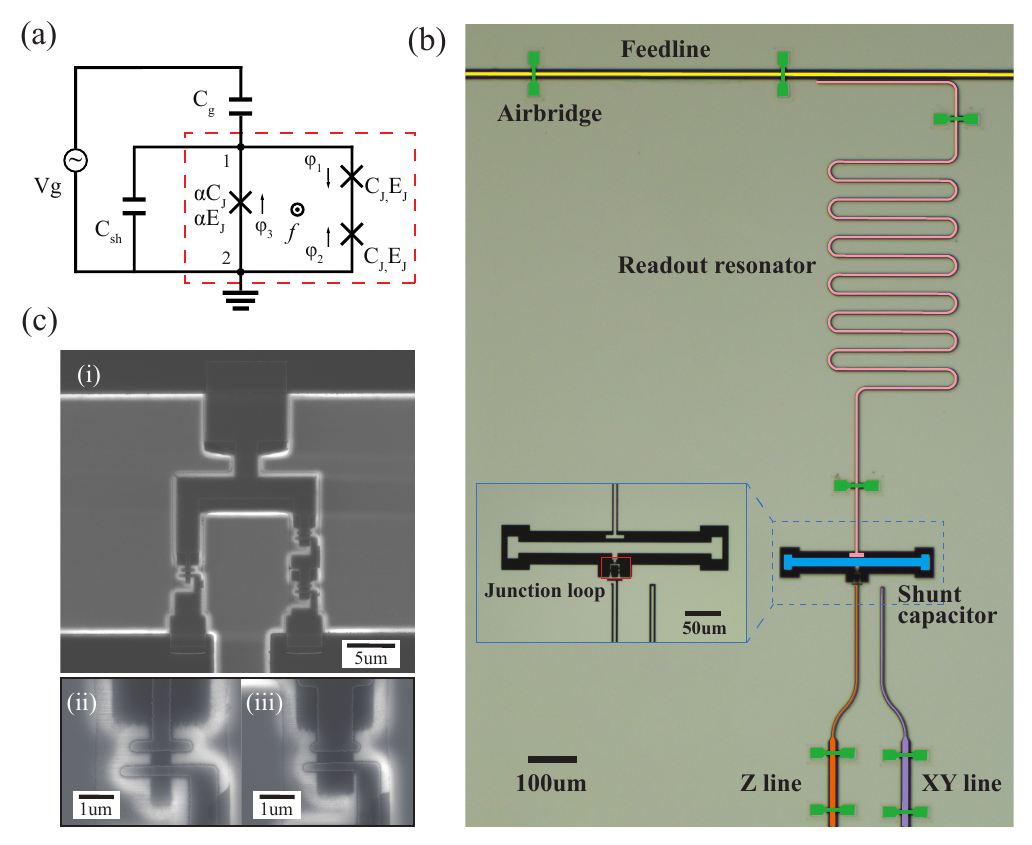}
    \caption{C-shunt flux qubit device. (a) Circuit schematic highlighting the Josephson junction loop (red dashed box). (b) False-color optical micrograph of the device, showing the feedline (yellow) coupled to a readout resonator (pink), the shunt capacitor (blue), airbridges (green) for ground connection, dedicated Z-bias (orange) and XY control (purple) lines for flux and microwave driving, respectively. The inset highlights the junction loop (red).  (c) Scanning electron micrographs of the junction area: (i) overview of the junction loop, (ii–iii) enlarged views of junctions with different sizes.} 
    \label{fig1}
    \end{figure}


\begin{figure*}[t]
    \centering
    \includegraphics[width=\textwidth]{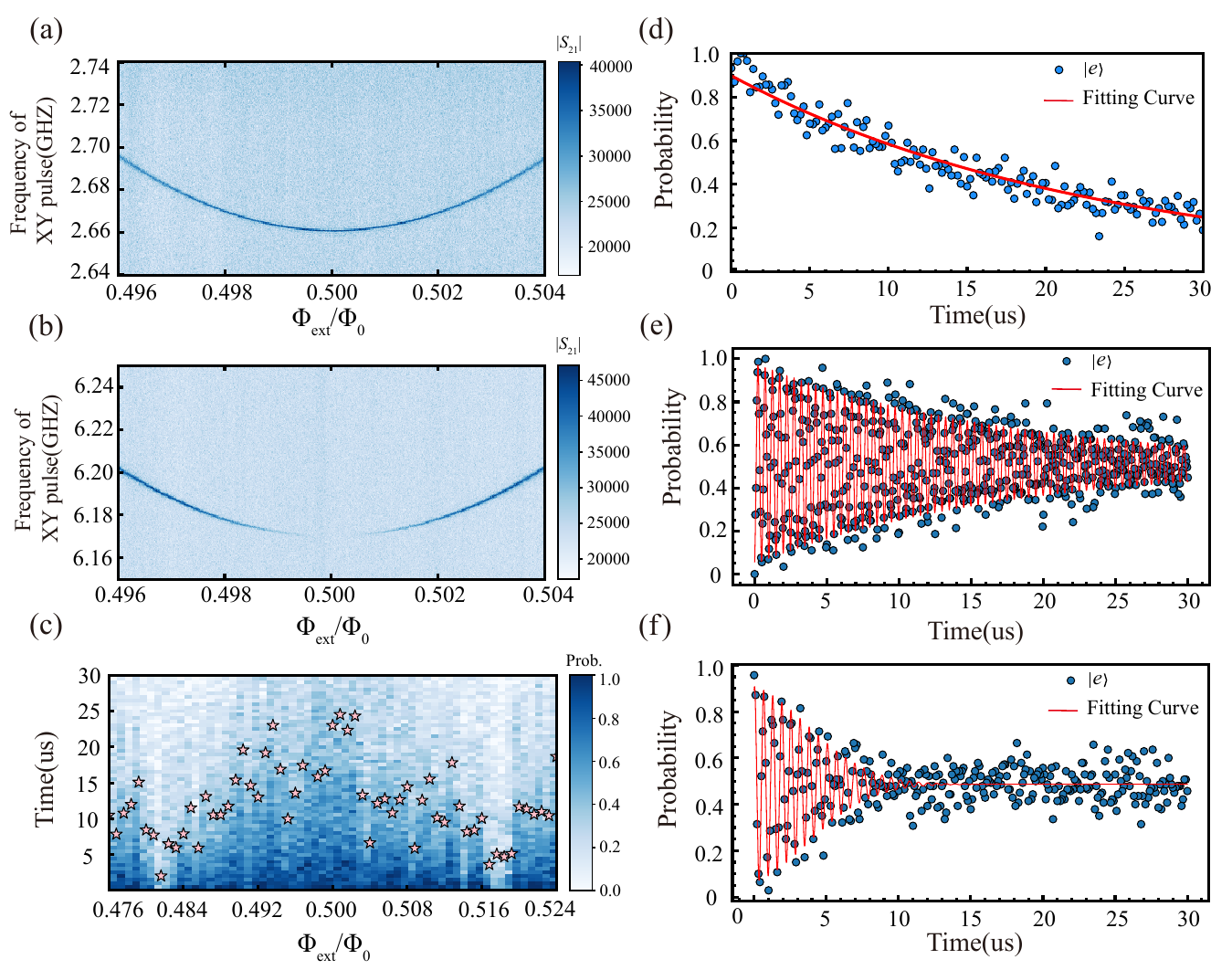}
    \caption{Performance characteristics of the qubit.(a-b) Qubit transition spectrum of the $\left|g\right\rangle \leftrightarrow |e\rangle$ and $|g\rangle \leftrightarrow |f\rangle$ transitions versus external magnetic flux. (c) Energy relaxation time $T_1$ versus external magnetic flux, showing a maximum around the sweet spot. (d–f) Representative measurements and exponential fits of relaxation and dephasing times at $\Phi_0/2$, yielding $T_1=23~\mu\mathrm{s}$, $T_2^{\mathrm{Ramsey}} = 6.3~\mu\mathrm{s}$, and $T_2^{\mathrm{spinecho}} = 17.4~\mu\mathrm{s}$ at $\Phi_0/2$, which indicate the coherence properties of the device.} 
    \label{fig2}
\end{figure*}

In this work, we present the design, fabrication, and experimental characterization of a C-shunt flux qubit with an emphasis on gate-level performance. We fabricated two devices with different shunt capacitances to explore the trade-off between anharmonicity and coherence~\cite{Yan2020}. As the shunt capacitance was reduced, the qubit anharmonicity increased from approximately $0.848~\mathrm{GHz}$ to $1.26~\mathrm{GHz}$, accompanied by a decrease in the relaxation time $T_{1}$ from $23~\mu\mathrm{s}$ to $6.6~\mu\mathrm{s}$ (detailed parameters are provided in the supplementary material). We selected the device with moderate anharmonicity ($848~\mathrm{MHz}$) and long relaxation time ($T_{1}=23~\mu\mathrm{s}$, $T_{2}^{\mathrm{spinecho}}=17.4~\mu\mathrm{s}$) as a balanced operating point for demonstrating high-fidelity single-qubit control. Using randomized benchmarking with optimized DRAG pulses, we achieve single-qubit gate fidelities exceeding $99.9\%$ for multiple gate operations. These results highlight that the C-shunt flux qubit can serve as a robust building block for high-fidelity quantum information processing.

The C-shunt flux qubit consists of a superconducting SQUID loop formed by three Josephson junctions and shunted with a large capacitance~\cite{yan2016}. Two of the junctions are identical, with Josephson energy $E_J$ and charging energy $E_C$, while the third junction is smaller, characterized by scaled parameters $\alpha E_J$ and $\alpha E_C$. The circuit schematic is shown in Fig.~\ref{fig1}{a}. The qubit Hamiltonian can be described using the standard flux-qubit model~\cite{Orlando1999}, where the external magnetic flux $\Phi_{\mathrm{ext}}$ controls the energy-level spacing and anharmonicity. The large shunt capacitance modifies the charging energy, reducing the qubit’s sensitivity to charge noise while maintaining strong anharmonicity. The Hamiltonian can be expressed as
\begin{align}
    H_{\text{csh}}=\frac{1}{2} \frac{P_{\text{p}}^{2}}{M_{\text{p}}^{2}} +\frac{1}{2} \frac{P_{\text{m}}^{2}}{M_{\text{m}}^{2}}+U\left(\varphi_{\mathrm{p}}, \varphi_{\mathrm{m}}\right),
\end{align}
where $P_{\text{p}} = -\mathrm{i} \partial / \partial \varphi_{\text{p}}$, 
    $P_{\text{m}} = -\mathrm{i} \partial / \partial \varphi_{\text{m}}$, $M_{\text{p}} = 2(\Phi_{\text{0}}/2\pi)^2C_{\text{J}}$, $M_{\text{m}} =2(\Phi_{\text{0}}/2\pi)^2C_{\text{J}}(1+2\alpha+2\beta)$ and the shunt capacitor is $C_{\text{sh}} = \beta C_{\text{J}}$. The effective potential is 
\begin{align}
U(\varphi_{\mathrm{p}}, \varphi_{\mathrm{m}}) = &  E_{\mathrm{J}}\{2(1-\cos \varphi_{\text{p}}\cos \varphi_{\text{m}}) \\ \nonumber
& +\alpha[1-\cos (2\pi f + 2\varphi_{\text{m}})]\}.  
\end{align}
The device was fabricated using nano-fabrication techniques~\cite{Martinis2002}, and the schematic is shown in Fig.~\ref{fig1}{b}. After pre-processing the substrate, a 100-nm-thick aluminum film was deposited via electron-beam evaporation. Large-scale structures, such as the resonator, shunt capacitor, and transmission lines, were patterned using laser direct writing and wet etching, while the junction area was defined by high-resolution electron-beam lithography followed by double-angle electron-beam evaporation~\cite{Dolan1977},which is crucial in determining the value of $\alpha$ (see Fig.~\ref{fig1}{c}). Airbridges were also fabricated to suppress slotline modes~\cite{Chen2014_Airbridges}. More detailed fabrication procedures are provided in supplementary material.

The performance of the device was characterized in a dilution refrigerator at a base temperature of 10 mk. The detailed room-temperature and cryogenic measurement setup follows our previous work~\cite{li2025}. The coherence properties of the C-shunt flux qubit were systematically characterized by spectroscopy and time-domain measurements, as summarized in Fig.~\ref{fig2}. Figures~\ref{fig2}(a) and \ref{fig2}(b) illustrate the flux-dependent transition spectrum, yielding resonance frequencies of $\omega_{ge}/2\pi = 2.661~\mathrm{GHz}$ and $\omega_{gf}/2\pi = 6.170~\mathrm{GHz}$ near the optimal bias point, from which we extract an anharmonicity $\mathcal{A}/2\pi=\omega_{ef}/2\pi-\omega_{ge}/2\pi=\omega_{gf}/2\pi-2\omega_{ge}/2\pi=848~\mathrm{MHz}$. Near the sweet spot ($\Phi_{\rm ext} \approx \Phi_0/2$), the qubit wave functions possess definite parity, which forbids direct $|g\rangle\!\leftrightarrow\!|f\rangle$ transitions. When the flux bias is detuned from $\Phi_0/2$, this symmetry is broken, and the $|g\rangle\!\leftrightarrow\!|f\rangle$ transition becomes allowed~\cite{liu2005,Vool2018,wang2024}. Such a sizable anharmonicity effectively suppresses leakage to higher levels, which in turn supports long coherence and the potential for high-fidelity gate operations~\cite{nguyen2019}. Figure~\ref{fig2}(c) displays the energy relaxation time $T_1$ as a function of the external magnetic flux, which exhibits a maximum value of $23~\mu\mathrm{s}$ at the sweet spot (see Fig.~\ref{fig2}(d)). Time-domain coherence experiments performed at $\Phi_{\mathrm{ext}} = \Phi_0/2$ yield $T_2^{\mathrm{Ramsey}} = 6.3~\mu\mathrm{s}$ and $T_2^{\mathrm{spinecho}} = 17.4~\mu\mathrm{s}$, as shown in Fig.~\ref{fig2}(e-f). The observed coherence properties of the C-shunt flux qubit are governed by the underlying device physics and dominant decoherence mechanisms, including flux noise and material-related losses. A detailed analysis of these effects is provided in supplementary material. Taken together, these metrics demonstrate that the C-shunt flux qubit, while maintaining strong anharmonicity, can also support long relaxation times, providing a robust platform for high-fidelity single-qubit control.


Building on this platform, we now describe the realization and characterization of single-qubit quantum gates. Single-qubit operations are implemented by applying resonant microwave drives to the qubit~\cite{Gambetta2011}. The system Hamiltonian in the lab frame can be written as
\begin{align}
H=\tfrac{1}{2}\hbar\omega_{ge}\sigma_z+\hbar\Omega(t)\cos(\omega_d t+\phi)\sigma_x,
\end{align}
where $\Omega(t)$ is the drive envelope, and $\phi$ is the drive phase. In the rotating frame under the rotating-wave approximation (RWA), the Hamiltonian reduces to
\begin{align}
H_{\text{drive}}=\tfrac{1}{2}\hbar\Omega(t)\big(\cos\phi\,\sigma_x+\sin\phi\,\sigma_y\big).
\end{align}
Thus, the drive phase $\phi$ determines the rotation axis on the Bloch sphere (e.g. $\phi=\pi/2$ for $Y$ rotations), while the pulse area $\theta=\int \Omega(t)\,dt$ sets the rotation angle (e.g. $\theta=\pi$ for an $X$ gate, $\theta=\pi/2$ for an $X/2$ or $Y/2$ gate)~\cite{Gustavsson2012}.

Owing to the strong anharmonicity of the C-shunt flux qubit, the energy separation between the computational subspace and higher excited states reaches approximately $848~\mathrm{MHz}$. Such a large anharmonicity provides sufficient spectral separation to support control pulses with durations on the nanosecond scale, for which the associated spectral bandwidth remains well below the anharmonicity, thereby suppressing leakage to non-computational states \cite{motzoi2009,Gambetta2011}. In practice, although such strong anharmonicity enables fast control in principle, the use of extremely short pulses places more stringent demands on the microwave control chain, including waveform distortion, finite bandwidth, and calibration accuracy. In this work, we therefore employ comparatively longer gate pulses to ensure robust and repeatable gate calibration and benchmarking.

In our experiment, we employ Gaussian pulses of 20~ns total length with a full width at half maximum (FWHM) of 10~ns~\cite{steffen2003,werninghaus2021a}. Although the C-shunt flux qubit features a relatively large anharmonicity of about 848~MHz, short pulses still lead to spectral broadening and spurious excitations to higher levels. To mitigate such leakage and phase errors induced by AC-Stark effect, we implement the Derivative Removal by Adiabatic Gate (DRAG) technique~\cite{motzoi2009,Gambetta2011}, in which a correction term proportional to the time derivative of the Gaussian envelope is applied in quadrature. The optimal DRAG coefficient $\eta$ and pulse amplitude are precisely calibrated using error-amplification sequences~\cite{chow2010,lucero2010} by repeating the target gates up to $m = 321$ times, ensuring the minimization of cumulative gate errors (see supplementary material Fig. S4 for detailed calibration curves).

To quantify the performance of our single-qubit gates, we employ randomized benchmarking (RB)~\cite{knill2008,magesan2011,Barends2014}. The decay of the ground-state survival probability in RB can be described by an exponential function,  
\begin{equation}
P(m) = A p^{m} + B
\end{equation}
where $m$ is the number of Clifford gates, $p$ is the depolarizing parameter, and $A$, $B$ account for state preparation and measurement (SPAM) errors. The average Clifford fidelity is given by $F_{\mathrm{Clifford}} = \tfrac{1+p}{2}$. To further extract the fidelity of a specific gate, we perform interleaved randomized benchmarking, where the target gate is interleaved between random Clifford operations. The corresponding decay is expressed as
\begin{equation}
P_{\mathrm{int}}(m) = A p_{\mathrm{int}}^{m} + B,
\end{equation}
with depolarizing parameter $p_{\mathrm{int}}$. The fidelity of the interleaved gate is calculated from the ratio of the decay parameters,
\begin{equation}
F_{\mathrm{gate}} = 1 - \frac{1 - p_{\mathrm{int}}/p}{2}.
\end{equation}

\begin{figure}[h]
    \centering
    \includegraphics[width=\columnwidth]{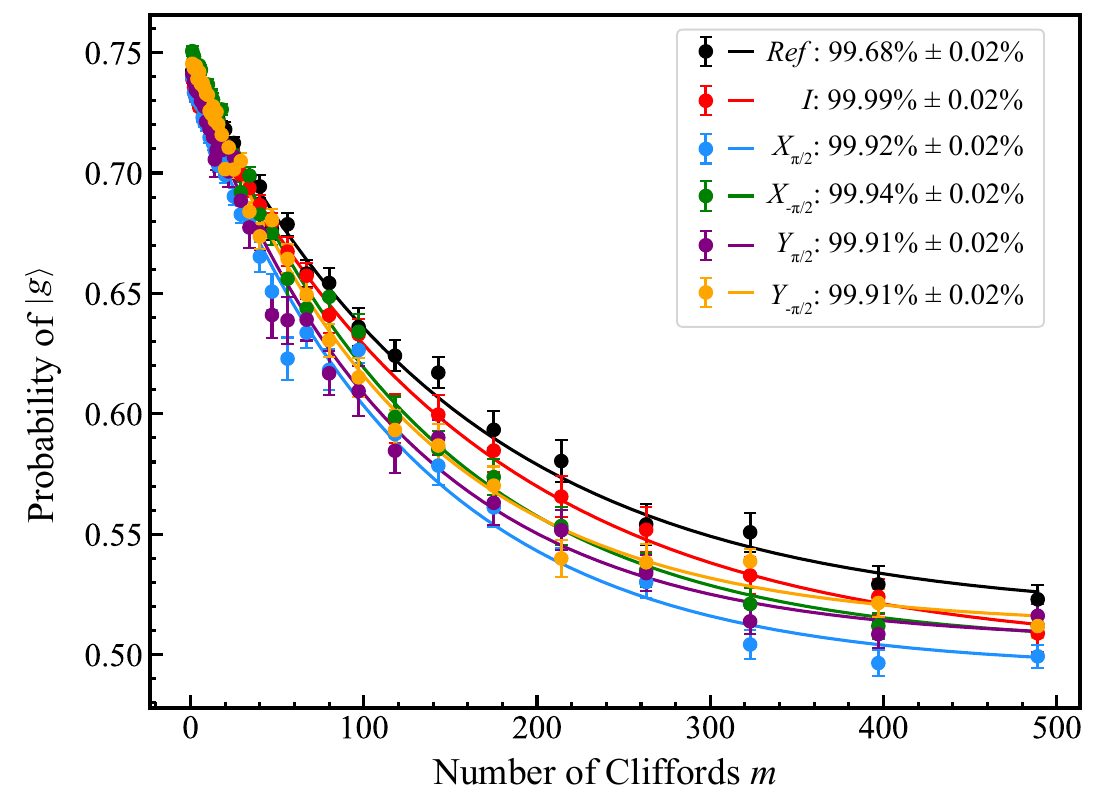}
    \caption{Randomized benchmarking (RB) of single-qubit gates on the C-shunt flux qubit.
    The plot shows the measured ground-state survival probability $P(|g\rangle)$ as a function of the number of Clifford gates. Data points represent experimental measurements with error bars from statistical uncertainty, while solid lines correspond to exponential fits. The initial survival probability matches the independently calibrated readout fidelity $F_e=0.75$ (see Fig. S6 in supplementary material). The extracted reference Clifford fidelity is $99.68\% \pm 0.02\%$. Single-gate fidelities for the identity gate and the $X_{\pm \pi/2}$ and $Y_{\pm \pi/2}$ rotations are obtained from interleaved randomized benchmarking using Eq.~(7). The resulting values are: identity gate ($I$, $99.99\% \pm 0.02\%$), $\pi/2$ and $-\pi/2$ rotations around $X$ ($99.92\% \pm 0.02\%$ and $99.94\% \pm 0.02\%$), and $\pi/2$ and $-\pi/2$ rotations around $Y$ ($99.91\% \pm 0.02\%$ and $99.91\% \pm 0.02\%$).}
    \label{fig3}
\end{figure}
By fitting the experimental data to these models, we extract the single-qubit gate fidelities. For each Clifford sequence length m, the ground-state survival probability was averaged over $k=50$ independently generated random Clifford sequences, each repeated for 4000 measurement shots. The quoted uncertainties of the extracted fidelities correspond to the standard errors obtained from the covariance matrix of the nonlinear least-squares fitting. Figure~\ref{fig3} displays the measured ground-state survival probability as a function of the number of Clifford gates. The black symbols correspond to the reference RB sequence, yielding an average Clifford fidelity of $99.68 \pm 0.02\%$. The colored symbols represent interleaved RB experiments for different single-qubit gates, with solid curves showing the corresponding exponential fits. From these fits, we extract gate fidelities of 
$99.99 \pm 0.02\%$ for the identity (I) gate, 
$99.92 \pm 0.02\%$ for the $X_{\pi/2}$ gate, 
$99.94 \pm 0.02\%$ for the $X_{-\pi/2}$ gate, 
$99.91 \pm 0.02\%$ for the $Y_{\pi/2}$ gate, and 
$99.91 \pm 0.02\%$ for the $-Y_{\pi/2}$ gate, respectively.
where the uncertainty ($\pm0.02\%$) reflects the statistical error from the exponential fitting. These results demonstrate that our single-qubit gate operations achieve fidelities well above the fault-tolerance threshold~\cite{Fowler2012,Terhal2015}. To further elucidate the origin of the observed gate fidelities, we present a separate error budget analysis in supplementary material, where the relative contributions from decoherence, control imperfections, and SPAM errors are discussed.

In summary, we have designed and fabricated a C-shunt flux qubit and systematically investigated its coherence properties and gate performance. The device exhibits an anharmonicity of about 848 MHz, an energy relaxation time $T_1$ of $23~\mu\text{s}$, and dephasing times $T_2^{Ramsey}$= 6.3 $\mu$s and $T_2^{spinecho}$ = 17.4 $\mu$s. Using randomized benchmarking, we demonstrated single-qubit gate fidelities exceeding 99.9\%. These results establish the C-shunt flux qubit as a robust platform for high-fidelity single-qubit control, combining strong anharmonicity with long coherence times in an experimentally accessible regime. In addition, recent theoretical and experimental studies indicate that C-shunt–based architectures can support engineered interactions and controlled ZZ coupling in multiqubit settings~\cite{Zhao2020,Ku2020,heunisch2023}, highlighting their potential for future scalable implementations.

See the supplementary material for detailed descriptions of the device fabrication, parameters, measurement procedures, as well as an analysis of the underlying device physics and detailed gate error budget.

 This work was supported by the Micro/Nano Fabrication Laboratory of Synergetic Extreme Condition User Facility (SECUF). The devices were made at the Nanofabrication Facilities at the Institute of Physics, CAS in Beijing. This work was supported by Quantum Science and Technology-National Science and Technology Major Project (Grant No. 2021ZD0301800), the National Natural Science Foundation of China (Grants No. 12574540, No. 92265207, No. T2121001, No. 92065112, No. 92365301, No. T2322030).

\bibliography{refrence}

\end{document}